\documentstyle[prd,aps,preprint]{revtex}
\begin{document}
\draft

%
%

\preprint{Nisho-96/4} \title{Ferromagnetic Domain Wall and 
Primordial Magnetic Field} \author{Aiichi Iwazaki}
\address{Department of Physics, Nishogakusha University, Shonan Ohi Chiba
  277,\ Japan.} \date{August 7, 1996} \maketitle
\begin{abstract}
We show that coherent magnetic field 
is generated spontaneously when a large domain
wall is created in the early universe. It is caused by 
two dimensional massless fermions bounded to the domain wall soliton.
We point out
that the magnetic field is a candidate of primordial magnetic field
\end{abstract}
\pacs{98.80.Cq, 95.85.Sz, 11.15.Tk, 11.30.Er\hspace*{3cm} 
  }
\vskip2pc


Domain walls\cite{kibble} arise in any unification models 
with a discrete symmetry
in Higgs potentials and the symmetry is often used to reduce complication in
the potentials. Unfortunately the domain walls are unfavorable in cosmological
point of view\cite{zeldovich}; energy density of the domain wall dominates 
in the universe and
invalidates the scenario of the history of the early universe. 
To cure this problem
several ideas\cite{zeldovich,preskill,rai,senjanovic} are proposed, 
but there is not yet a definite solution.
Here we do not address this problem although we suggest 
a possible solution for the problem throughout analysis 
of new features of the domain wall.


 In this paper we analyze
 magnetic properties
of the domain wall ( hereafter we simply call it as the wall ) and 
derive a favorable consequence of an origin of 
primordial magnetic field\cite{hogan} in the universe. 
We show spontaneous generation 
of a magnetic field in the presence of the wall. 
A size of the coherence of the magnetic field is comparable to 
that of the wall.
 ( We assume the existence of a domain wall with its size $L$ being  
the same as the distance to the horizon, as indicated by 
numerical simulation\cite{vachaspani} 
of the creation of
the walls in the early universe ).
The essence of the result is in the presence of fermion 
zero modes\cite{jakiew}
on the wall; the zero modes exist in general when fermion couples with
 the domain wall soliton. It implies the existence of 
massless fermions on the two dimensional wall, although the fermions are
 massive when they are unbounded to the wall. Such two dimensional
massless fermions have a specific magnetic property 
when they have electric charges; their magnetic moments
are infinite ( we may define the magnetic moment such as a derivative
of fermion energy in terms of magnetic field evaluated 
at the value of the vanishing field ).
This fact suggests that the energy of the fermion gas ( 
 two dimensional fermion gas confined to the wall ) 
may has unusual dependence
of magnetic field $B$ perpendicular to the wall; usually 
the energy is proportional to $B^2$. 
As we will show later,
 the free energy of the fermion gas
is proportional to $B$. The sign of 
the coefficient of this term is negative. 
Thus the amount of the decrease of the energy 
is proportional to $B$. On the other hand 
the field energy is proportional
to $B^2$.
Therefore, spontaneous generation of magnetic field can occur since
total energy of the system decreases with the presence of the sufficiently
weak magnetic field.
 As the size of the wall is the same as the distance to the horizon, 
the scale of the coherence of such magnetic field
is so large that it may be a candidate of primordial magnetic field
leading to intergalactic or galactic magnetic field\cite{parker}.
 The strength of the field 
depends on the periods when the wall is created and when it disappears. 
For instance, if it is created in the electroweak phase transition at about 
the temperature of $1$ TeV, the strength of the field generated at the 
temperature is about $10^{9}$ Gauss. It leads to the field with its strength
$10^{-15}$ Gauss at recombination of photons and electrons.
We suggest that since the generation 
of the magnetic field breaks explicitly CP invariance,
it might lead to a possible solution for the domain wall problem
in a model with spontaneous CP violation.

We now present detail calculations. Hereafter we consider 
a large flat domain wall with its size $L$ assumed to be 
the same as the distance to the horizon. 
First we discuss fermionic zero modes\cite{jakiew}. 
The modes arise in general when
the fermion $\psi$ couples with a Higgs field $\phi$; strictly speaking
the existence of the modes depends on how the field $\phi$ couples
with the fermion. Here we
assume for simplicity the real Higgs field
 $\phi$ whose potential has a discrete symmetry, $\phi \to -\phi$ 
and the following Yukawa coupling, $ g\bar{\psi}\psi\phi + h.c.$. 
The field $\phi$ behaves near the wall supposed to be
sited at $x_3=0$ such as $\phi(x_3) \to \pm v$ 
as $ x_3 \to \pm \infty$; $\pm v$
are vacuum expectation values of the Higgs field $\phi$. The wall is extending 
in $x_1$ and $x_2$ directions.
 Then, it is easy to find the zero modes by solving a Dirac equation with 
energy $E$ and derivatives $\vec{\partial}$ transverse to the wall put 
to be zero, $(E\gamma_0 + i\vec{\gamma}\vec{\partial} + i\gamma_3\partial_3
   + g\phi)\psi = 0$ 
where $\vec{\gamma}$ $(\vec{\partial})$ is a two dimensional vector. Adopting
the representation of gamma metrics,

\begin{equation}
\gamma_0 = \left(
 \begin{array}{cc}
       1 & 0 \\
       0 & -1
  \end{array}
    \right) \ \  {\rm and} \ \ 
\gamma^i =  \left( 
  \begin{array}{cc}
          0 & \sigma_i \\
  -\sigma_i & 0
\end{array}
\right) 
\end{equation}  
we find the solutions with zero energy,

\begin{equation}
 \psi = {u\choose i\sigma_3u}, \quad
 u = {a\choose b} exp(-z(x_3)) \\
\label{bound}
\end{equation} 
where $z(x_3)$ is defined as $\partial z(x_3)/\partial x_3 = \phi(x_3)$; 
$a$ and $b$ are constants. Thus there are two zero energy modes.
 These may be viewed as zero energy bound states of massive fermion $\psi $ 
with its mass  
 $m=\sqrt{g} v$. 
It turns out that when the states 
move in transverse directions on the wall 
( in this case $a$ and $b$ are functions of $x_1$ and $x_2$ ),
their energies are proportional to
their momenta. Thus they are massless fermions in
two dimensional space of the wall. Note that such states attached to
the wall loose $2$ dynamical degrees of freedom; components $u_1$ and
$u_2$ of $\psi = {u_1\choose u_2}$ are not independent with each others.
Namely spin degrees of freedom are lost so that the two
dimensional massless fermion ( anti-fermion ) has only one dynamical
degree of freedom\cite{jackiew}.

In order to show spontaneous generation of magnetic field 
we calculate a thermodynamical potential and 
a zero point energy of these fermions
under a magnetic field perpendicular to the wall. The fermions are
 supposed to have electric charge, $e$.
For the purpose  we need to find energy spectrum under a constant magnetic 
field $B$; $\vec{B} = ( 0, 0, B )$. Choosing a gauge
such as $\vec{A} = ( -x_2, x_1, 0 ) B / 2$, we solve the above 
Dirac equation with gauge potential $\vec{A}$. Then,
 relevant solutions take the form in eq(\ref{bound}) with both of a and b
being functions of $x_1$ and $x_2$ ; these are
solutions of the cyclotron motions of the fermions bounded to the wall. 
Their spectra can be obtained easily, 
$E_n = \pm\sqrt{2eBn}$ ( $n\geq 0$ ) with degeneracy per unit erea being 
given by $eB/2\pi$.
Eigenstates are characterized with integer $n$ and orbital 
angular momentum $m$ ( $n, m\geq 0$ )

\begin{equation}
\psi = {u\choose i{\sigma}_3 u}, \quad u = c {v_1\choose v_2} 
\label{psi}
\end{equation}
with
\begin{equation}
v_1 = \rho^m e^{im\theta}{L_m}^{(n)}e^{-eB\rho^2/4},\ \
{\rm and},\ \ v_2 = \rho^m e^{i(m+1)\theta}\partial_{\rho}{L_m}^{(n)}
e^{-eB\rho^2/4}/E_n
\label{v}
\end{equation}
where ${L_m}^{(n)}(eB\rho^2/2)$ is Laguerre function 
and $c$ is a normalization constant ( $\rho^2=x_1^2+x_2^2$ ).

It should be mentioned that these states representing fermions 
bounded to the wall need to satisfy
a energy condition such as $E_n\leq m$. Otherwise, the states represent 
fermions unbounded to the wall and hence are not relevant to the property
of the wall. This energy condition is distinctive of the fermions on the wall.
Besides those solutions, there are solutions describing states of 
cyclotron motions in $x_1-x_2$ plane and scattering states
in $x_3$ direction. These are also not relevant for
intrinsic properties of the wall.


Strictly speaking, there are energy eigenstates localized on the wall
with their energies higher than m. But it seems that these states are
unstable against any couplings with scattering states with the same energies.
They decay and corresponding fermions go away from the wall. Hence 
it is conceived that they are irrelevant for the property of the wall.


Now, let us calculate thermodynamical potential $\Omega$ 
under the condition that 
the temperature is much smaller than the mass of the fermion.
Then it is given for 
free gas of fermions
 such that

\begin{equation}
\label{potent}
\Omega = \Omega_{+} + \Omega_{-}+ \Omega_0, \quad  \Omega_{\pm,}=
-\beta^{-1}N_d\sum_{n=1}^{\infty}\log{(1+e^{-E_n\beta})}, \quad \Omega_0=
-\beta^{-1}N_d\log{2} 
\end{equation}
where $N_d=eBL^2/2\pi$ is degeneracy of a Landau level ( $L^2$ is surface area
of the wall ). 
Summation $\sum_{E_n\leq m}$ resulted from the energy condition 
has been replaced by the summation $\sum_{n=1}^{\infty}$
owing to the condition, $\beta^{-1}<<m$. The indices 
( $\pm,0$ ) denote quantities of fermions, anti-fermions and
zero energy states ( $E_{n=0}=0$ ), respectively. 
$\beta$ is inverse temperature.
For simplicity we have assumed a vanishing chemical potential associated with 
the fermion number, namely
the fermion number in the universe is assumed to be negligibly 
small just as baryon number in the universe.

As we are interested in the behavior of the system at small magnetic field,
the sum over integer $n$ can be evaluated 
by using Poisson resummation formula,

\begin{equation}
\sum_{n=1}^{\infty}f(n\hat{\beta}^2)=
\sum_{m=-\infty}^{\infty}\int_{0}^{\infty}dx f(x\hat{\beta}^2)e^{2\pi imx}=
\int_{0}^{\infty}dx f(x\hat{\beta}^2)+2\sum_{m=1}^{\infty}\int_{0}^{\infty}
dx f(x\hat{\beta}^2)cos(2\pi mx)
\end{equation}
with $\hat{\beta}^2 \equiv 2eB\beta^2$, where $f(x)=\log(1+e^{-\sqrt{x}})$.
Expanding the second term in small $\hat{\beta}$,
we obtain the thermodynamical potential for $\hat{\beta}\to 0$ such that

\begin{equation}
\Omega=-\frac 1{2\pi}\beta^{-3}L^2\int_0^{\infty}\frac{y^2dy}{1+e^y}
+\frac{\sqrt{2eB}eB\zeta(-1/2)L^2}{2\pi} + O(\beta (eB)^2)
\label{omega}
\end{equation}
where the first term represents a contribution of free massless fermions 
without the magnetic field and the second one represents a correction by
the weak magnetic field $eB<<\beta^{-2}$. This first term
 can be also obtained by using the free fermion energy,
 $E_k=\left|k \right|$
( $k$ is $2$-dimensional momentum ) and by taking account of both of fermion
and anti-fermion with only one dynamical degree of freedom.

Similarly we calculate a zero point energy of the two dimensional massless
fermion on the wall. In the case we need to take care the energy condition; 
energies of these fermions are restricted such that $E_n\leq m$.
This is because the fermions with higher energies 
than their three dimensional mass $m=gv$ are not 
bounded to the domain wall. Thus zero point oscillations of such fermions 
are not associated with the property of the wall, but with 
the property of the outside of 
the wall. Therefore the zero point energy of the wall is given by 

\begin{equation}
E_0=-\sum_{n=1}^F \sqrt{2eBn}N_d \approx -\frac{m^3L^2}{6\pi}
-\frac{eBmL^2}{4\pi}-\frac{\sqrt{2eB}eB\zeta(-1/2)L^2}{2\pi}
\end{equation} 
where $F=m^2/2eB$ is the number of Landau levels, energies of whose
states
are less than the mass $m$ of the fermion; $\zeta(z)$ is zeta-function.
 We have taken a limit of small
magnetic field ( $eB<<m^2$ ) in the calculation. 
The first term represents a zero point 
energy of the fermion without magnetic field. The term can be 
also expressed
by the integration, $-L^2\int_{|k|<m}|k|dk^2/(2\pi)^2$. This quantity
should be renormalized to surface energy of the wall. The third term has 
already been obtained \cite{hosotani} as a zero point energy 
of two dimensional massless fermion
without the energy condition mentioned above.

Comment is in order. In the calculation of the zero point energy we have 
included only effects of localized states whose energies are less than $m$. 
Namely,
we have not taken into account of the effects of states localized on the wall
with higher energies than $m$.  
Such states really exist in the case of no their coupings with 
the scattering states. The states, however, are quite unstable against 
the coupings; these localized states decay and disappear. Hence it is 
reasonable to conceive that the only states 
which really exist are scattering states with energies higher than $m$ and 
bound states with energies less than $m$.  Thus we expect physically that 
in a realistic circumstance we may neglect the effects of 
the localized states with higher energies than $m$. 
On this point we elaborate it in future publication.

We can see that the magnetic field
reduces both the thermal energy of the fermion gas and the zero point energy; 
the amounts of the reduction are
proportional to $B\sqrt{B}L^2$ and $BL^2$ respectively. 
This fact leads to spontaneous generation
of magnetic field. Namely when the magnetic field is present in the universe
the whole free energy $E_B$ associated with the magnetic field is given by
the above energies of two dimensional fermion gas and the energy of
the magnetic field itself,

\begin{equation}
E_B=-\frac{eBmL^2}{4\pi} + \frac{B^2 L^3}{2\mu}
\label{a}
\end{equation}  
where $L^3$ is the volume inside the horizon. 
$\mu$ is permeability of the universe. 
( The thermal effect of the order of $B^{3/2}$ in eq(\ref{omega}) cancels with
the corresponding effect of the vacuum energy\cite{Voloshin}. But this 
cancellation 
does not necessarily hold in any cases: For instance 
when the chemical potencial of 
the fermion in the universe is nonvanishing, the cancellation does not occur. 
In the case the thermal effect dominates over the effect 
of the vacuum energy.) 
Hence, the magnetic field generated spontaneously is obtained 
by minimizing this energy,

\begin{equation}
B_r = \frac{em\mu}{4\pi L}
\label{magne}
\end{equation}
This is the magnetic field associated with the domain wall soliton 
in the universe. The strength of 
the field becomes small with the expansion of the universe.
 In the radiation-dominated 
universe the distance $L$ to the horizon behaves 
with temperature $\beta^{-1}$ such as $f M_{pl}\beta^2$ 
( $M_{pl}$ is the Planck mass and $f$ is the total number of massless 
degrees of freedom at temperature $\beta^{-1}$ ) so that 
$B_r$ decreases in such a way as
 $B_r \sim \beta^{-2}/f M_{pl}$. Numerically, when the domain wall arises
at the electroweak phase transition, e.g. $\beta^{-1}\sim 100$ GeV ( $f\sim
100$ ), 
the magnitude
of the magnetic field is of order of $10^6$ Gauss 
with the use of the top quark mass $m=175GeV$.

Here we wish to understand 
a mechanism of the above phenomena ( ferromagnetism ) of the wall. 
For the purpose we calculate magnetization, 
$M = -\partial E_B /\partial B|_{B\to 0} = emL^2/4\pi$. 
It turns out that the magnetization originates in the zero point energy.
Namely, each state with negative energy $E'_n=-\sqrt{2eBn}$ contributes a 
positively infinite magnetic moment, while the number 
$N_d$ ( $=eBL^2/2\pi$ ) of the states with 
the energy vanishes as $B\to 0$ and the number $F$ ( $=m^2/2eB$ ) 
of the Landau levels in 
the Dirac sea becomes infinite as $B\to 0$. These effects leads to 
the spontaneous generation of the magnetic moment of the wall.
This is a mechanism leading to
the ferromagnetism of the wall in the case of 
the sufficiently small temperature.

We have so far assumed in the evaluation of the thermodynamical potential
that the temperature is much smaller than the mass of the fermion, i.e.
$\beta^{-1} << m$. The fermions bounded to the wall never
escape even if they are excited thermally: So we have performed infinite 
sum in the previous calculation. On the contrary when
the temperature is much larger than the mass of the fermion,
they can escape from the wall. In the evaluation of the thermodynamical
potential in the case we should take account of 
the energy condition; only the fermions with
their energies less than the mass $m$ contribute to the potential,

\begin{equation}
 \Omega_{\pm,}=
-\beta^{-1}N_d\sum_{n=1}^{F}\log{(1+e^{-E_n\beta})}
\cong -\beta^{-1}N_d F \log{2},\quad \Omega_0=
-\beta^{-1}N_d\log{2}
\end{equation}
where we have taken the temperature $\beta^{-1}$ being much larger than
the mass $m$ of the fermion.
Therefore the whole free energy depending on the magnetic field is

\begin{equation}
E_B=-\frac{eBL^2\log{2}}{2\pi\beta} +\frac{B^2L^3}{2\mu}
\label{b}
\end{equation}
where the zero point energy has been neglected because $\beta^{-1}$ is much 
larger than $m$ so that the energy is much smaller than the thermal energy 
of the first term in eq(\ref{b}). 
The thermal energy dominates over the zero point energy.
We see that imposition of the magnetic field reduces the free energy of the 
fermion gas.
Then the field generated spontaneously is given by

\begin{equation}
B_r=\frac{e\mu\log{2}}{2\pi\beta L}
\label{br}
\end{equation}
Numerically, $B_r\cong 10^6$ Gauss at $\beta^{-1}=100$ GeV, or $B_r\cong 1$
Gauss at $\beta^{-1}=1$ GeV.

Comparing this result with the previous one in eq(\ref{magne}), we understand
that the magnetic field $B_r$ originates in the zero point oscillation of 
the fermion when the temperature is less than the mass of the fermion, 
while the field originates in the thermal effects of the fermion when the 
temperature is higher than the mass. In the latter case the magnetic 
field increases
the number $N_{B}$ of the states with energies less than the mass $m$,

\begin{equation}
N_{B}-N_{B=0}=\frac{eBL^2}{2\pi}\sum_{n=0}^{F-1}-
 \frac{L^2}{(2\pi)^2}\int_{|k|\leq m}dk^2=
\frac{eBL^2(F-m^2/2eB)}{2\pi}>0
\end{equation}
where $F$ is such that $\sqrt{2eBF}>m>\sqrt{2eB(F-1)}$.
Thus it leads to increases entropy $S$ of the gas and hence decreases the free
energy ( $=E-S\beta^{-1}\approx -S\beta^{-1}$ ). 
This is the reason of the spontaneous generation of the magnetic field
when the temperature is higher than the mass of the fermion 
( $E\approx mL^2 << S\beta^{-1}$ ).

As we have shown, the spontaneous generation of magnetic field occurs 
in the presence of the domain wall soliton. The origin of the magnetic field 
is a natural consequence of fermion dynamics on the domain wall soliton.
Hereafter we wish to discuss briefly 
phenomenological application of the result.

Let us consider a realistic model where
the wall is created at the electroweak phase transition and 
the fermions are quarks or lepton: Such models exist, e.g. a next to minimal
supersymmetric standard model\cite{Abel}. The transition temperature may be 
about $1$ TeV
and it is much larger than any masses of quarks and leptons. Thus the above
formula in eq(\ref{br}) is applicable for this case.
Then, the magnetic field
is roughly $10^{9}$ Gauss 
at the temperature of $1$ TeV with its coherent length $L=10^{-2}$cm.
This magnetic field is naively expected to lead to
a magnetic field of
$10^{-15}$ Gauss with coherent length $10^{10}$ cm around 
the period of the recombination. Note that the evolution of the field is 
such that $Ba^2$ = constant because its flux is conserved 
owing to very large electric conductivity of the universe\cite{parker}; $a$ is 
the cosmic scale factor in Robertson-Walker metric. 
As the universe expands, the size of the wall becomes large and the magnetic 
field becomes weak. But, we expect that 
a mechanism\cite{zeldovich,preskill,rai,senjanovic} begins
to work for destabilizing the wall before the wall dominates the energy
density of the universe. Then the wall decays with its shape topologically 
equivalent to sphere. In the case the strength of the magnetic field 
is turned out to be proportional 
to the inverse of radius of the sphere. Hence the magnetic field becomes 
stronger as the spherical wall shrinks more. 
In these processes of the wall
evolution the magnetic fields are printed in medium of the universe because
their fluxes are conserved.
Hence the magnetic fields with various strengths and sizes 
of coherence are left in the evolution of the wall. These are candidates of 
primordial magnetic fields leading to galactic or stellar magnetic field.
To determine the strength ( the coherent length ) of 
the field more precisely, we have
to include dissipation of magnetic field 
and the number of the fermion species, 
and to know when and how the wall 
disappears.

Finally we point out a possible solution for the domain wall problem in a 
electroweak model with spontaneous CP violation. In the model domain walls 
may arise associated with
the spontaneous breakdown of this discrete CP invariance.
If quarks or leptons couple with 
Higgs fields of the wall and give rise to zero modes, 
then the magnetic field may be generated 
spontaneously in a similar way as we have discussed. This field is not 
CP invariant. Thus an effective Higgs potential
may involve a CP non invariant term which would be generated by fermion loops
with the effect of this magnetic field. The term induces 
a vacuum pressure\cite{zeldovich}
which makes eventually the wall to disappear before 
it dominates in the universe.
A problem in this scenario 
is whether or not this CP non invariant term gives rise to the vacuum pressure
strong enough to solve
the domain wall problem. These possibilities are now under investigation.


In summary, we have shown that the magnetic field is generated spontaneously
when the domain wall is present in the early
universe. This magnetic field is a candidate of 
a primodial magnetic field leading
to intergalactic or galactic magnetic field in the present
universe. We have only discussed the flat domain wall in this paper, but
similar results can be obtained even in the case of spherical domain walls. 
We will present these results in future publications.

The author would like to thank Prof. Y. Hosotani for indicating importance of 
the zero point energy and valuable comments. He also thank Prof. M. Kawasaki 
for useful discussions and 
staff members of theory division in
Institute for Nuclear Study, University of Tokyo for their
hospitality.




\end{document}